\pdfoutput=1
\documentclass[runningheads]{llncs}

\usepackage[T1]{fontenc}

\usepackage{graphicx}

\PassOptionsToPackage{hyphens}{url}\usepackage[hidelinks]{hyperref}
\usepackage{caption}
\usepackage{placeins}
\hypersetup{
  colorlinks = true,
  linkcolor  = blue,
  urlcolor   = blue,
  citecolor  = black,
}
\newcommand*{\fullref}[1]{\hyperref[{#1}]{\autoref*{#1} \nameref*{#1}}}
\newcommand*{\secref}[1]{\hyperref[{#1}]{\autoref*{#1}}}
\usepackage{svg}
\usepackage{enumitem}
\newlist{inlineenum}{enumerate*}{1}
\setlist*[inlineenum]{mode=unboxed,label=(\arabic*)}
\newlist{inlineitem}{itemize*}{1}
\setlist*[inlineitem]{label=\textbullet}
\usepackage{multirow,tabularx, makecell, booktabs}
\newcolumntype{Y}{>{\centering\arraybackslash}X}

\begin{document}
\title{SWARM-SLR - Streamlined Workflow Automation for Machine-actionable Systematic Literature Reviews}
\titlerunning{SWARM-SLR: Workflow Automation for SLR}
\author{Tim Wittenborg\inst{1, 2}\orcidID{0009-0000-9933-8922} \and
Oliver Karras\inst{3}\orcidID{0000-0001-5336-6899} \and
Sören Auer\inst{1,2,3}\orcidID{0000-0002-0698-2864}}
\authorrunning{Wittenborg et al.}
%
\institute{L3S Research Center, Leibniz University Hannover, Hannover, Germany\\
\email{\{tim.wittenborg, soeren.auer\}@l3s.de} \and Cluster of Excellence SE²A – Sustainable and Energy-Efficient Aviation, Technische Universität
Braunschweig, Germany \and
TIB - Leibniz Information Centre for Science and Technology, Hannover, Germany\\
\email{\{oliver.karras, soeren.auer\}@tib.eu}}
\maketitle

\begin{abstract}
Authoring survey or review articles still requires significant tedious manual effort, despite many advancements in research knowledge management having the potential to improve efficiency, reproducibility, and reuse.
However, these advancements bring forth an increasing number of approaches, tools, and systems, which often cover only specific stages and lack a comprehensive workflow utilizing their task-specific strengths.
We propose the Streamlined Workflow Automation for Machine-actionable Systematic Literature Reviews (SWARM-SLR) to crowdsource the improvement of SLR efficiency while maintaining scientific integrity in a state-of-the-art knowledge discovery and distribution process.
The workflow aims to domain-independently support researchers in collaboratively and sustainably managing the rising scholarly knowledge corpus.
By synthesizing guidelines from the literature, we have composed a set of 65 requirements, spanning from planning to reporting a review.
Existing tools were assessed against these requirements and synthesized into the SWARM-SLR workflow prototype, a ready-for-operation software support tool.
The SWARM-SLR was evaluated via two online surveys, which largely confirmed the validity of the 65 requirements and situated 11 tools to the different life-cycle stages.
The SWARM-SLR workflow was similarly evaluated and found to be supporting almost the entire span of an SLR, excelling specifically in search and retrieval, information extraction, knowledge synthesis, and distribution.
Our SWARM-SLR requirements and workflow support tool streamlines the SLR support for researchers, allowing sustainable collaboration by linking individual efficiency improvements to crowdsourced knowledge management.
If these efforts are continued, we expect the increasing number of tools to be manageable and usable inside fully structured, (semi-)automated literature review workflows.

\end{abstract}

\section{Introduction}
A key pillar of the scientific method is the synthesis of various research contributions in survey or review articles.
The number of reviews claiming to be ``comprehensive'', however, declined over the past two decades with the number of relevant publications rapidly increasing~\cite{larsen_understanding_2018}.
Larsen et al.~\cite{larsen_understanding_2018} show that the current practice of authoring surveys does not scale to this increase in workload, is error-prone, and lacks reproducibility and reuse.
Even in fields like medicine, where systematic protocols were established for meta-analysis, authors like Ioanidis~\cite{Ioannidis2016TheMP} conclude that there is mass production of redundant, misleading, and conflicting systematic reviews and meta-analyses.

Systematic Review Automation (SRA) aims to utilize "computational assistance and automation to dramatically improve the process" while being "faster, with fewer resources"~\cite{SLR-20}.
With software solutions becoming more accessible while data machine-readability increases, the future of scholarly knowledge discovery and distribution can be greatly improved~\cite{jaradeh_open_2019}.
Evidently, this productivity acceleration can reduce the required time from months to only two weeks, while apparently maintaining or even increasing scientific integrity~\cite{clark_full_2020}.
Artificial Intelligence (AI) further promises to automate repetitive tasks and support others in so-called AI-based literature reviews (AILRs)~\cite{wagner_artificial_2022}.

These advancements face reservations and doubts, as they further highlight the requirement of "human interpretation and insightful syntheses, as well as novel explanation and theory building."~\cite[p. 13]{wagner_artificial_2022}.
With Large Language Models (LLM) promising a new interface to digital libraries, information retrieval, and knowledge synthesis become more accessible to a broader user group.
The Texas University Library specifically lists ChatGPT as an ``AI-Based Literature Review Tool''~\cite{xiao_research_nodate}.
These user groups may struggle to maintain their domain integrity when faced with unreliable results caused by the inherent hallucination of LLMs and misunderstood processes~\cite{Kathryn_chatgpt_2023}.
A transparent capability assessment is required, educating on capabilities and limits, specifically when using AI with hallucinating capabilities, where the user is not to be perceived as the author, but a commissioner~\cite{noauthor_copyright_2023}.

This inflation of problems, solutions, and requirements warrants an accessible and efficient literature review workflow~\cite{shabanov_effortless_2023}, specifying which tasks to solve through using which tools to satisfy defined requirements.
When facing an exponentially growing scholarly knowledge corpus~\cite{bornmann_growth_2015}, an SLR workflow should enable researchers across domains to collaboratively structure knowledge Findable, Accessible, Interoperable, and Reusable (FAIR)~\cite{wilkinson_fair_2016}.
With this work, we provide insights into how software tools can be used to systematically reduce inconsistency by streamlining an accessible literature review process toward crowd- or expert-sourced knowledge management.

Our contributions include:
(1) We synthesize SLR guidelines into 65 requirements for the \textbf{S}treamlined \textbf{W}orkflow \textbf{A}utomation fo\textbf{r} \textbf{M}achine-actionable \textbf{S}ystematic \textbf{L}iterature \textbf{R}eviews (\textit{SWARM-SLR requirements}), allowing distributed collaboration and development of cohesive and comprehensive SRA.
(2) We propose the ready-to-use \textit{SWARM-SLR workflow} spanning interoperable SRA from planning to reporting a review, demonstrating the usage of software tools situated according to their strengths and weaknesses (\textit{SWARM-SLR tools}).

To achieve this, we define the scope of SWARM-SLR in \secref{sec:background}.
With that groundwork laid out, we refine the requirement and toolsets in \secref{sec:approach} to synthesize them into a workflow.
The requirements, tools, and workflow are evaluated in \secref{sec:evaluation} and discussed in \secref{sec:discussion}, allowing us to draw a promising conclusion about the future of SWARM-SLR in \secref{sec:conclusion}.

\section{Related Work\label{sec:background}}
We consider related work regarding \nameref{sub:LR} and \nameref{sub:ST} in their significance for \nameref{sub:AILR}.

\subsubsection{Systematic Literature Reviews (SLR)\label{sub:LR}}
aim to identify, analyze, synthesize, and report literature~\cite{torraco_writing_2016}.
There are many categorizations of different literature review types~\cite{adams_research_2007,ralph_paving_2022}.
According to Adams et al. \cite{adams_research_2007}, review types may be categorized into ``Instrumental Review'', ``Evaluative Review'' and ``Exploratory Review'', of which SWARM-SLR aims to improve primarily the latter.
The individual approach varies depending on domain, context, and goal, which is why our approach focuses on answering specific research questions, one of five purposes of a literature survey according to Torraco~\cite{torraco_writing_2016}.
The author also states that ``the body of literature on literature reviews has grown from a few publications to a large body of literature'', covering often interchangeably used terms like ``literature review'', ``integrative literature review'', ``research review'' and ``review of the literature'', while the term ``review'' is often substituted for ``survey''.
For the context of this paper, SWARM-SLR is domain-independently targeted at researchers conducting either a literature survey or review, especially when conducting an exploratory review answering specific research questions.

\subsubsection{Software Tools\label{sub:ST}}
of various degrees of reliability, particularly with the general advances of AI, have been introduced into scientific processes~\cite{shabanov_effortless_2023}, surfacing novel challenges in AI literacy education to maintain integrity~\cite{liu_summary_2023}.
Especially since tools like Large Language Models (LLM) ``display wider scope and autonomy'',~\cite[p. 5]{novelli_taking_2023} there is a wide array of misconceptions about what these tools can do and where the user has to supervise the results themselves.
While there are risks inherent to handing off responsibility to these tools, other dedicated survey support software has been developed, increasing the efficiency while maintaining integrity.
Open Knowledge Maps (OKMaps)~\cite{kraker_peter_2016_4705327,Open_Knowledge_Maps_2019} specifically addresses accessible user engagement in search processes and focuses on reliable discovery.
Similar web search assistant tools like Connected Papers~\cite{noauthor_connected_nodate} and long-established services like Google Scholar or Semantic Scholar also address these problems~\cite{liu_co-citation_2022,Mart_n_Mart_n_2018}.
Many of these tools integrate with one another. ArXiv~\cite{noauthor_arxivorg_nodate} for example has various references and embeds Connected Papers on its website, which again builds on Semantic Scholar.
This redundancy allows users to easily verify information retrieved at one location with other sources.
Meta services like the Open Research Knowledge Graph (ORKG)~\cite{stocker_fair_2023} can make use of this traceable knowledge journey as a core tool in the literature review process, especially in the final steps.
The Best Practices in Data Science and Artificial Intelligence by Borisova et al.~\cite{borisova_open_2023} specifically recommend using the ORKG, especially in combination with semantic annotation through SciKGTeX~\cite{bless2023scikgtex}.
This specialization of dedicated tools, capable of reliably solving certain tasks, stands in contrast to the general capabilities of recent conversational agents without certainty in any area.
This leads to an incoherent spread of tools across the workflow.
To situate each tool where it excels, we need to understand what work needs to be done in the first place.

\subsubsection{Systematic Review Automation (SRA)\label{sub:AILR}}
is an ongoing topic in scientific discourse~\cite{wallace_modernizing_2013,SLR-23,SLR-20,kohl_online_2018,marshall_toward_2019,wagner_artificial_2022}.
Particularly AI-based Literature Reviews (AILR) are highlighted for their potential by Wagner et al.~\cite{wagner_artificial_2022}.
Wagner et. al., Kohl et al.~\cite{kohl_online_2018}, as well as Pulsiri and Vatananan-Thesenvitz~\cite{pulsiri_improving_2018} all formulate different frameworks for assessing different capabilities of SLR tools, naming different tasks and evaluating the fitness of tools differently.
Pulsiri and Vatananan-Thesenvitz for example created an overview on automating the SLR.
They adapt the structure illustrated in \secref{table:Pulsiri} to analyze how and where the state-of-the-art SLR can be automated.

\begin{table}[h]\centering
    \vspace{-0.55cm}
    \captionsetup{width=\linewidth}
    \captionof{table}{Mapping papers in relation to automation methods and SLR according to Pulsiri and Vatananan-Thesenvitz~\cite{pulsiri_improving_2018}.}
    \label{table:Pulsiri}
        \small
        \begin{tabular}{c|c|c|c}\toprule
            \bf Stage & \bf Task & \bf Semi-automated & \bf Fully automated\\\midrule
             \multirow{2}{*}{I} & \it Planning a review & \cite{SLR-20}\cite{SLR-21}\cite{SLR-23} & \\
             & \it Defining a scope & \cite{SLR-20}\cite{SLR-21}\cite{SLR-23} & \\\hline
             \multirow{3}{*}{II} & \it Search & \cite{SLR-15}\cite{SLR-16}\cite{SLR-21} & \cite{SLR-19}\cite{SLR-23}\cite{SLR-27}\\
             & \it Select & \cite{SLR-14}\cite{SLR-15}\cite{SLR-16}\cite{SLR-21}\cite{SLR-25}\cite{SLR-26}\cite{SLR-28} & \cite{SLR-19}\cite{SLR-20}\cite{SLR-23}\\
             & \it Evaluate & \cite{SLR-16}\cite{SLR-21} & \cite{SLR-19}\cite{SLR-20}\cite{SLR-23}\\\hline
             \multirow{2}{*}{III} & \it Analysis & \cite{SLR-16}\cite{SLR-20}\cite{SLR-21}\cite{SLR-22}\cite{SLR-23} & \\
             & \it Synthesis & \cite{SLR-16}\cite{SLR-20}\cite{SLR-21}\cite{SLR-23} & \\\hline
             \multirow{1}{*}{IV} & \it Reporting & \cite{SLR-16}\cite{SLR-20}\cite{SLR-21}\cite{SLR-23} & \\
             \bottomrule
        \end{tabular}
    \normalfont
    \vspace{-0.35cm}
\end{table}

This table has the benefit of being tailored to tools supporting the review process, similar to online registers such as ``Systematic Review Toolbox'' with over 100 tools, 78 of which apply to the systematic literature review~\cite{noauthor_systematic_nodate}.
These tools, however, are mostly not standalone applications that can be readily applied to any scientist's workflow. 
Such tools have been studied by Bosman and Kramer in their extensive work of collecting over 700 tools at the time of the writing of this work~\cite{noauthor_400_tools_and_innovations_in_scholarly_communication_nodate,bosman_innovations_2016}.
While not mapped directly to the review tasks displayed in \tablename~\ref{table:Pulsiri}, they are categorized into 30 phases of scholarly communication, many of which can be situated inside the literature survey process.
A selection of these tools has also been classified according to their fitness for 17 specific tasks, along with a dedicated survey with 20,000 respondents solidifying this mapping~\cite{noauthor_tools_nodate}.
Similarly, research-paper recommender systems were reviewed by Beel et al.~\cite{beel_research-paper_2016} and Kreutz and Schenkel~\cite{kreutz_scientific_2022}. 
In total, they provide an overview of over 150 paper recommender systems, as well as two examples of how SLR might be conducted.
All of these tools provide a strong set to choose from when designing a framework for the SWARM-SLR while highlighting the need to streamline a cohesive workflow to navigate them efficiently.

We have now established the elusive definition of the literature review process as well as the sheer abundance of tools addressing their interpretation and scope.
What they lack is a consistent, fundamental framework of what tasks and steps an SLR requires. 
Since currently no tool can achieve SRA alone, their strengths and weaknesses need to be comparable over the entire scope.
Given this plethora of options, it is very likely that task-excelling tools combined into a modular workflow would make SRA more achievable than any current tool available.

\section{Approach\label{sec:approach}}
\begin{figure}[h]
\vspace{-0.25cm}
    \centering
    \includegraphics[width=.9\textwidth]{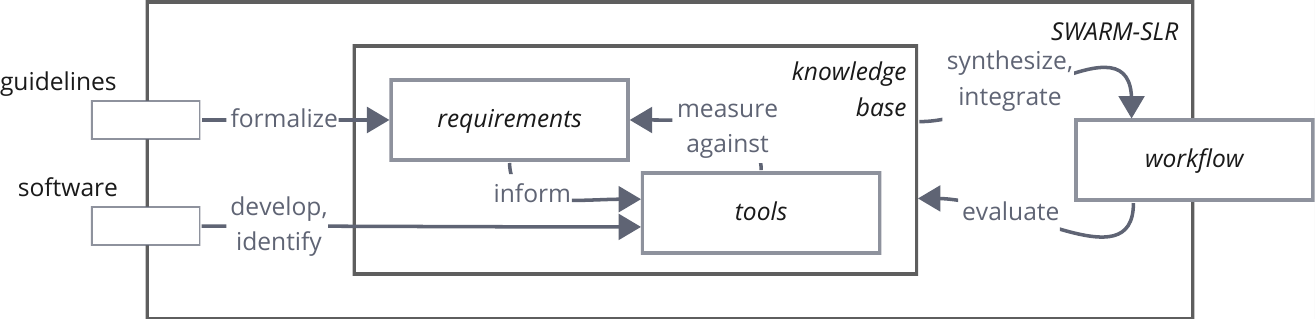}
    \caption{Formalizing SWARM-SLR requirements informs the identification and development of the tool catalog. Synthesizing these into a tool-assisted workflow makes this knowledge actionable and enables evaluation.}
    \label{fig:approach}
\vspace{-0.65cm}
\end{figure}

To properly describe the approach described in \secref{fig:approach}, we need to define \textit{SWARM-SLR requirements}, \textit{SWARM-SLR tools} and \textit{SWARM-SLR workflow}.

\paragraph{Requirements.}
To realize streamlined workflow automation, it is advisable to formalize requirements. Requirements enable the comparison of tools and workflows by allowing us to assess them against a common set of objectives.
These objectives may be structured into groups and should formally represent the stages and tasks of a systematic literature review.
In this context, \textit{SWARM-SLR requirements} describe a set of requirements formalized from SLR literature, specifically with the intent to assess the applicability of \textit{SWARM-SLR tools}.

\paragraph{Tools.}
A tool enables solving a given task at a higher efficiency or effectiveness. In the context of scientific workflow automation, specifically for literature reviews, tools are usually software tools, but categorically they can be of any nature.
While chaining software tools into a toolchain is easier due to interoperability, building a tool-assisted workflow implementing non-software physical tools is also possible.
In this context, \textit{SWARM-SLR tools} describe a set of tools measured against the \textit{SWARM-SLR requirements}, with the potential to be used as a tool catalog for the \textit{SWARM-SLR workflow}.

\paragraph{Workflow.}
A workflow structures required operations alongside tools and recommendations to achieve a higher efficiency or effectiveness.
In this context, the \textit{SWARM-SLR workflow} describes guidance alongside the \textit{SWARM-SLR requirements} while implementing and interconnecting \textit{SWARM-SLR tools}.

\begin{figure}[h]
    \centering
    \includegraphics[width=.6\textwidth]{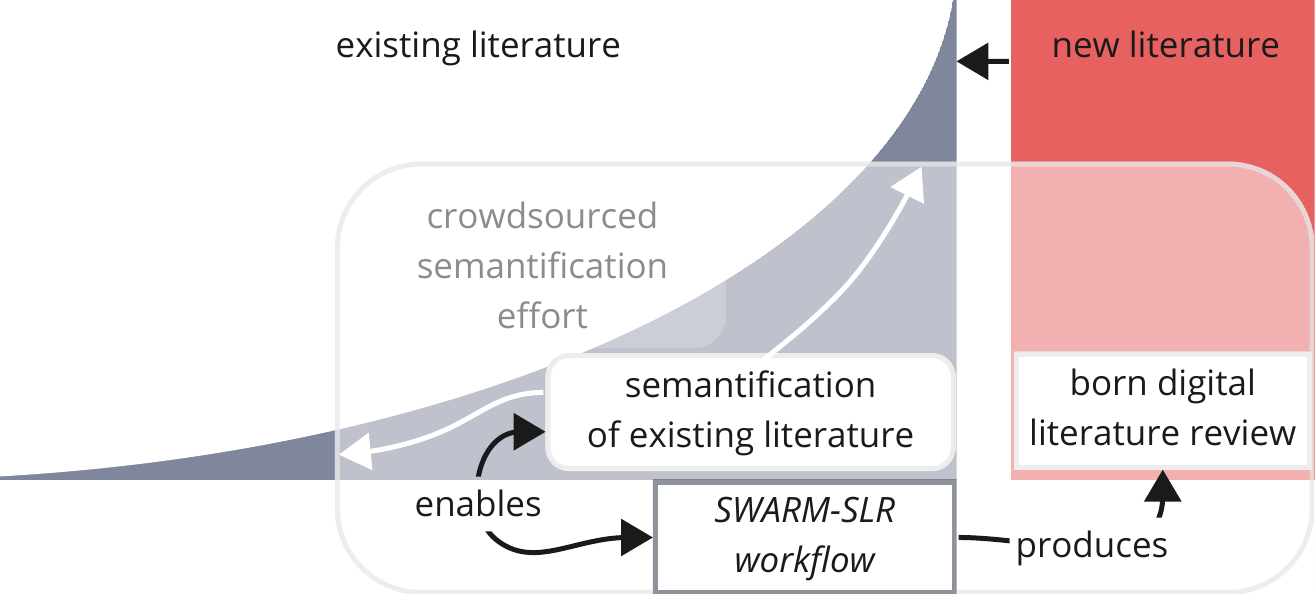}
    \caption{The SWARM-SLR workflow enables and is enabled by the semantification of literature. It also produces a born-digital literature review, catalyzing future knowledge synthesis.}
    \label{fig:application}
\vspace{-0.25cm}
\end{figure}

With an efficient SWARM-SLR workflow, future SLR have the potential to scale against the increasing amount of publications each year~\cite{bornmann_growth_2015} depicted in \secref{fig:application}.
By integrating the semantification of existing literature in a collaborative environment, each SWARM-SLR review potentially contributes to one another, reducing the workload of future research regarding pre-existing literature.

The approach was iteratively developed following the design science methodology using a demo survey case of aerospace engineering knowledge representations.
This enabled us to refine certain requirements and test a first selection of tools.
Simultaneously, these explorations allowed us to integrate additional requirements and tools, either arising by the process itself or suggested from the surveyed literature, for example, the Natural Language Processing (NLP) pipeline by Berquand~\cite{phdthesisBerquand}.
The respective development process and result are presented in the subsections \nameref{sec:req}, \nameref{sec:tools} and \nameref{sec:workflow}.

\subsection{Requirements\label{sec:req}}
Using the baseline of Pulsiri and Vatananan-Thesenvitz, we used guidance from Wagner, Lukyanenko, Templier and Paré \cite{wagner_artificial_2022,templier_transparency_2018}, particularly for the first three stages (see \secref{table:Pulsiri}).
We detailed each task into individual steps, results, and requirements with basic literature like the Cochrane Handbook~\cite{higgins_cochrane_2023} and especially Machi and McEvoy~\cite{machi2012literature}, a shared source of all the aforementioned authors.

This version and its resulting requirements were iteratively discussed among authors and fellow scholars, especially from the fields of NLP and digital libraries.
Once the coverage was optimized to the best of the authors' knowledge, these requirements were properly formalized according to the guideline by Rupp~\cite{rupp_requirements-engineering_2014}. In the end, we managed to detail the eight tasks from \secref{table:Pulsiri} into 65 specific requirements, as shown in \secref{table:req1} to \secref{table:req3}.

\begin{table}[h]\centering
\vspace{-0.45cm}
\captionof{table}{SWARM-SLR requirements of stage I.}
\label{table:req1}
\scriptsize
\begin{tabularx}{\linewidth}{clccX}
\toprule
\textbf{Task} &\textbf{Step} &\textbf{Result} &\textbf{ID} &\textbf{Requirement:} The tool should be able to help the user to efficiently ... for the <result>. \\\midrule
\multirowcell{2}{\bf 1. Planning\\\bf a review} &\multirowcell{2}[0pt][l]{\it 1. Formulate\\ \it research interest} &\multirowcell{2}{explicit informal\\research interest} & 1 & formalize abstract interest \\ \cline{4-5}
& & & 2 & utilize central keywords \\ \midrule
\multirowcell{13}{\bf 2. Defining\\\bf a scope} &\multirowcell{3}[0pt][l]{\it 1. Check for related\\ \it research questions} &\multirowcell{3}{narrowed informal\\research interest} & 3 & know what previous scholarly works already covered \\ \cline{4-5}
& & & 4 & know what previous scholarly works were missing \\ \cline{2-5}
&\multirowcell{3}[0pt][l]{\it 2. Refine\\ \it scientific interest} &\multirowcell{3}{specific\\research question} & 5 & use specific vocabulary \\ \cline{4-5}
& & & 6 & state a clear scope \\ \cline{4-5}
& & & 7 & state a clear perspective \\ \cline{2-5}
&\multirowcell{2}[0pt][l]{\it 3. Formulate\\ \it Search Query} &\multirowcell{2}{preliminary weighted\\keywords and queries} & 8 & use relevant keywords \\ \cline{4-5}
& & & 9 & weight individual keywords \\ \cline{2-5}
&\multirowcell{2}[0pt][l]{\it 4. Refine with\\ \it related literature} &\multirowcell{2}{weighted keywords and\\refined queries} & 10 & extend the vocabulary \\ \cline{4-5}
& & & 11 & identify polysemes and synonyms \\ \cline{2-5}
&\multirowcell{4}[0pt][l]{\it 5. Reevaluate with\\ \it domain experts} &\multirowcell{4}{validated weighted\\keywords and queries} & 12 & validate research questions \\ \cline{4-5}
& & & 13 & validate keywords \\ \cline{4-5}
& & & 14 & validate weights \\ \cline{4-5}
& & & 15 & validate search query \\
\bottomrule
\end{tabularx}
\normalfont
\end{table}

\begin{table}[h]\centering
\vspace{-0.45cm}
\captionof{table}{SWARM-SLR requirements of stage II}
\label{table:req2}
\small
\scriptsize
\begin{tabularx}{\linewidth}{clccX}\toprule
\textbf{Task} &\textbf{Step} &\textbf{Result} &\textbf{ID} &\textbf{Requirement:} The tool should be able to help the user to efficiently ... for the <result>. \\
\midrule
\multirowcell{6}{\bf 3.\\\bf Search} &\multirowcell{3}[0pt][l]{\it 1. Find resources} &\multirowcell{3}{preliminary\\document set} & 16 &use reliable sources \\ \cline{4-5}
& & & 17 &find relevant documents \\ \cline{4-5}
& & & 18 &find similar documents \\ \cline{2-5}
&\multirowcell{3}[0pt][l]{\it 2. Remove duplicates\\\it and find\\\it missing documents} &\multirowcell{3}{curated\\document set} & 19 &identify duplicate documents \\ \cline{4-5}
& & & 20 &find unindexed documents \\ \cline{4-5}
& & & 21 &identify document set gaps \\ \midrule
\multirowcell{21}{\bf 4.\\\bf Select} &\multirowcell{13}[0pt][l]{\it 1. Extract\\\it structured data\\\it from documents} &\multirowcell{4}{metadata} & 22 &extract publication date \\ \cline{4-5}
& & & 23 &extract author(s) \\ \cline{4-5}
& & & 24 &extract publication venue \\ \cline{4-5}
& & & 25 &extract specified keywords \\ \cline{3-5}
& &\multirowcell{2}{content} & 26 &extract document text \\ \cline{4-5}
& & & 27 &differentiate chapters \\ \cline{3-5}
& &\multirowcell{6}{bag of words} & 28 &remove special character \\ \cline{4-5}
& & & 29 &identify multi-words \\ \cline{4-5}
& & & 30 &expand acronyms \\ \cline{4-5}
& & & 31 &lemmatise words \\ \cline{4-5}
& & & 32 &normalize words \\ \cline{4-5}
& & & 33 &remove stopwords \\ \cline{3-5}
& &contribution statements & 34 &identify statements \\ \cline{2-5}
&\multirowcell{8}[0pt][l]{\it 2. Calculate relational\\\it measurements within\\\it the document set} &\multirowcell{4}{document\\representation} & 35 &calculate ``term frequency'' (tf) and ``term frequency - inverse document frequency'' (tf-idf) \\ \cline{4-5}
& & & 36 &calculate word embedding and document embedding \\ \cline{4-5}
& & & 37 &represent document machine-readable \\ \cline{3-5}
& &\multirowcell{3}{similarity of\\documents within\\the document set} & 38 &consider synonyms \\ \cline{4-5}
& & & 39 &consider polysemes \\ \cline{4-5}
& & & 40 &calculate document similarity \\ \cline{2-5}
&\multirowcell{2}[0pt][l]{\it 3. Identify documents\\\it relevant for the\\\it research questions} &\multirowcell{2}{relevance of\\documents for\\research question} & 41 &represent research question machine-readable \\ \cline{4-5}
& & & 42 &calculate document relevancy for research question \\ \midrule
\multirowcell{6}{\bf 5.\\\bf Eval\\\bf -uate} &\multirowcell{2}[0pt][l]{\it 1. Remove out-of-scope\\\it document subset} &\multirowcell{2}{in-scope\\document set} & 43 &define in-/exclusion criteria \\ \cline{4-5}
& & & 44 &exclude documents \\ \cline{2-5}
&\multirowcell{2}[0pt][l]{\it 2. Evaluate\\\it relevancy measurement} &\multirowcell{2}{selection approval} & 45 &evaluate similarity of documents \\ \cline{4-5}
& & & 46 &evaluate relevance of documents \\ \cline{2-5}
&\multirowcell{2}[0pt][l]{\it 3. Classify\\\it document subsets} &\multirowcell{2}{literature subsets} & 47 &calculate variables \\ \cline{4-5}
& & & 48 & divide document set into subsets \\
\bottomrule
\end{tabularx}
\normalfont
\vspace{-0.45cm}
\end{table}

\newcommand{\dbottomrule}{\specialrule{0.3pt}{0pt}{0.4pt}
            \specialrule{1pt}{0pt}{\belowrulesep}
            }

\begin{table}[h]\centering
\vspace{-0.45cm}
\captionof{table}{SWARM-SLR requirements of stage III and IV}
\label{table:req3}
\small
\scriptsize
\begin{tabularx}{\linewidth}{ccccX}\toprule
\textbf{Task} &\textbf{Step} &\textbf{Result} &\textbf{ID} &\makecell[X]{\textbf{Requirement:} The tool should be able to help the user to efficiently ...\\for the <result>.} \\\midrule
\multirowcell{3}{\bf 6.\\\bf Analysis} &\multirowcell{3}[0pt][l]{\it 1. Annotate\\\it resources} &\multirowcell{3}{contribution\\representation} & 49 &annotate the document \\ \cline{4-5}
& & & 50 &represent the annotation machine-readable \\ \cline{4-5}
& & & 51 &annotate collaboratively \\ \midrule
\multirowcell{9}{\bf 7.\\\bf Synthesis} &\multirowcell{2}[0pt][l]{\it 1. Compare\\\it contributions} &\multirowcell{2}{comparison} & 52 &define comparison properties \\ \cline{4-5}
& & & 53 &insert contribution values \\ \cline{2-5}
&\multirowcell{2}[0pt][l]{\it 2. Craft arguments\\\it based on findings} &\multirowcell{2}{document claims} & 54 &identify evidence (data) \\ \cline{4-5}
& & & 55 &establish warrant \\ \cline{2-5}
&\multirowcell{5}[0pt][l]{\it 3. Critique\\\it the literature} &\multirowcell{5}{thesis claims} & 56 &identify evidence (claims) \\ \cline{4-5}
& & & 57 &identify patterns \\ \cline{4-5}
& & & 58 &detect fallacies \\ \cline{4-5}
& & & 59 &identify conflicts \\ \cline{4-5}
& & & 60 &resolve conflicts \\
\midrule
\multirowcell{5}{\bf 8.\\\bf Reporting} &\multirowcell{5}[0pt][l]{\it 1. Write the\\\it literature review} &\multirowcell{5}{review article} & 61 &utilise comparisons \\ \cline{4-5}
& & & 62 &support claims \\ \cline{4-5}
& & & 63 &establish outline \\ \cline{4-5}
& & & 64 &draft iteration(s) \\ \cline{4-5}
& & & 65 &audit iteration(s) \\ 
\bottomrule
\end{tabularx}
\normalfont
\end{table}

\subsection{Tools\label{sec:tools}}
Given the requirement set, the next task was to identify the most suitable tools. Using the references from \fullref{sub:AILR}, we composed a catalog including but not limited to the following tools: CADIMA; Large Language Models (LLM) like ChatGPT, Llama; search and recommendation engines like Google Scholar, Semantic Scholar, Open Knowledge Maps (OKMaps), ConnectedPapers; Zotero; Python; Natural Language Processing libraries like NLTK, spaCy, FLAIR; PDF miners like pdfminer, pypdf2; LaTeX; Open Research Knowledge Graph (ORKG); SciKGTeX; Colandr; Cochrane RevMan.

\begin{figure}[bt]
    \centering
    \includegraphics[width=\textwidth]{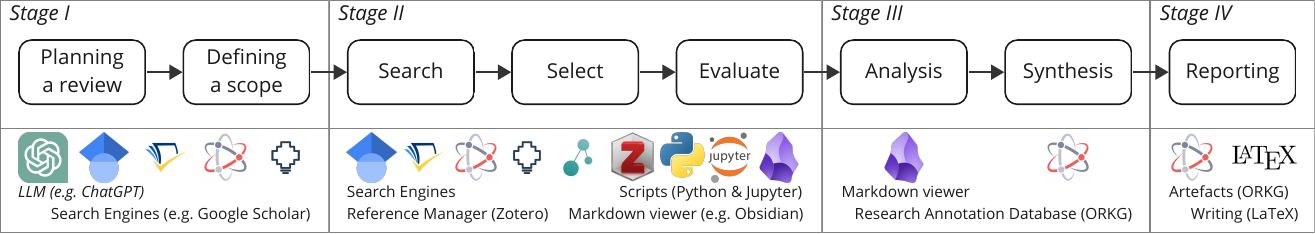}
    \caption{SWARM-SLR workflow guided by four online Jupyter Notebooks.}
    \label{fig:workflow}
\vspace{-0.45cm}
\end{figure}

As the set grew, an internal evaluation based on the requirements, known tool compatibility, user community, and previous experience with the tools resulted in a preliminary selection of the first 13 tools listed above to begin workflow implementation as depicted in \secref{fig:workflow}.

While \textit{CADIMA} was of exceptional interest as it structures the SLR, its rigid structure set it on the back of the list, providing a guideline similar to the requirements, but not an implementation priority by itself. 
Using \textit{Large Language Models (LLM) like ChatGPT or Llama}, however, was exceptionally incentivized due to their rising popularity, simultaneously leveraging new potential as well as providing SWARM-SLR users evidence and education regarding LLM limitations.
\textit{Google Scholar} and \textit{Semantic Scholar} were included due to their exceptional search capability, while \textit{Open Knowledge Maps (OKMaps)} and \textit{ConnectedPapers} complimented them without significantly increasing the overhead.
\textit{Zotero} was chosen for reference management due to being free and widely used in the community, providing an interface from search results to a local library.
\textit{Python}, implementing \textit{NLP libraries like NLTK, spaCy, FLAIR} and \textit{PDF miners like pdfminer, pypdf2}, was used to analyze the document set.
While using \textit{Open Research Knowledge Graph (ORKG)} alongside \textit{SciKGTeX} was recommended by the Best Practices in Data Science and Artificial Intelligence~\cite{borisova_open_2023}, the ORKG seemed particularly suitable for storing both intermediate and final products of the workflow.
For publication and formatting, SciKGTeX leads to the selection of \textit{LaTeX}, amongst other interoperability benefits.
Despite not being listed specifically as SWARM-SLR tools, supplementary tools like a coding IDE (\textit{Visual Studio Code}), a markdown viewer (\textit{Obsidian}), or PDF viewer (\textit{Acrobat Reader}) will be used by most users.

\subsection{Workflow\label{sec:workflow}}
The workflow is composed of four \nameref{par:script} connecting the tools using a Python \nameref{par:package}. 
Printing the GitHub representation of all notebooks, including examples, code snippets, LLM prompts, and replies, as well as illustrations, would comprise over 25 pages.
We refer to the repository for details.\footnote{\label{note:github-SWARM-SLR}\url{https://github.com/borgnetzwerk/tools/tree/main/scripts/SWARM-SLR}}

\begin{figure}[tb]
    \centering
    \includegraphics[width=\textwidth]{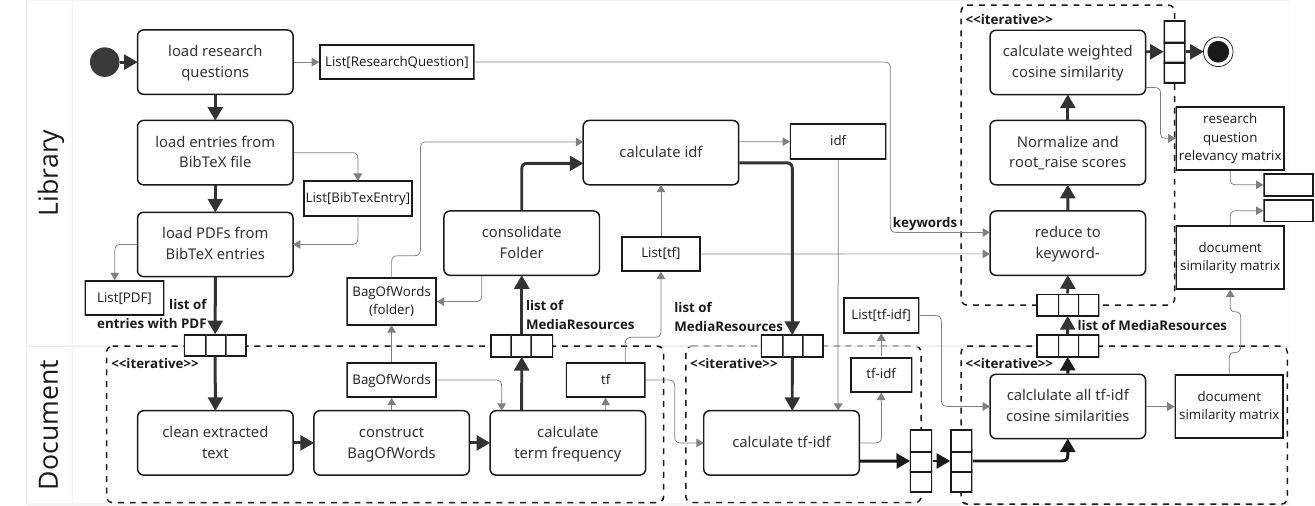}
    \caption{UML activity diagram for the fully automated \textit{Task 4: Select}}
    \label{fig:task4d}
    \vspace{-0.45cm}
\end{figure}

\paragraph{Package.\label{par:package}}
To automate certain tasks, we built and released the open source python package bnw\textunderscore{}tools~\cite{borgnetzwerk_borgnetzwerktools_2024}.
The package implements tool interfaces, analyzing Zotero entries to produce insights in Obsidian readable files, using PDF miners and NLP.
As \secref{fig:task4d} illustrates, it can automate certain tasks like\break{}\textit{Task 4: Select} when it is installed and executed in the local library. 

\begin{figure}[tb]
    \vspace{-0.45cm}
    \centering
    \includegraphics[width=\textwidth]{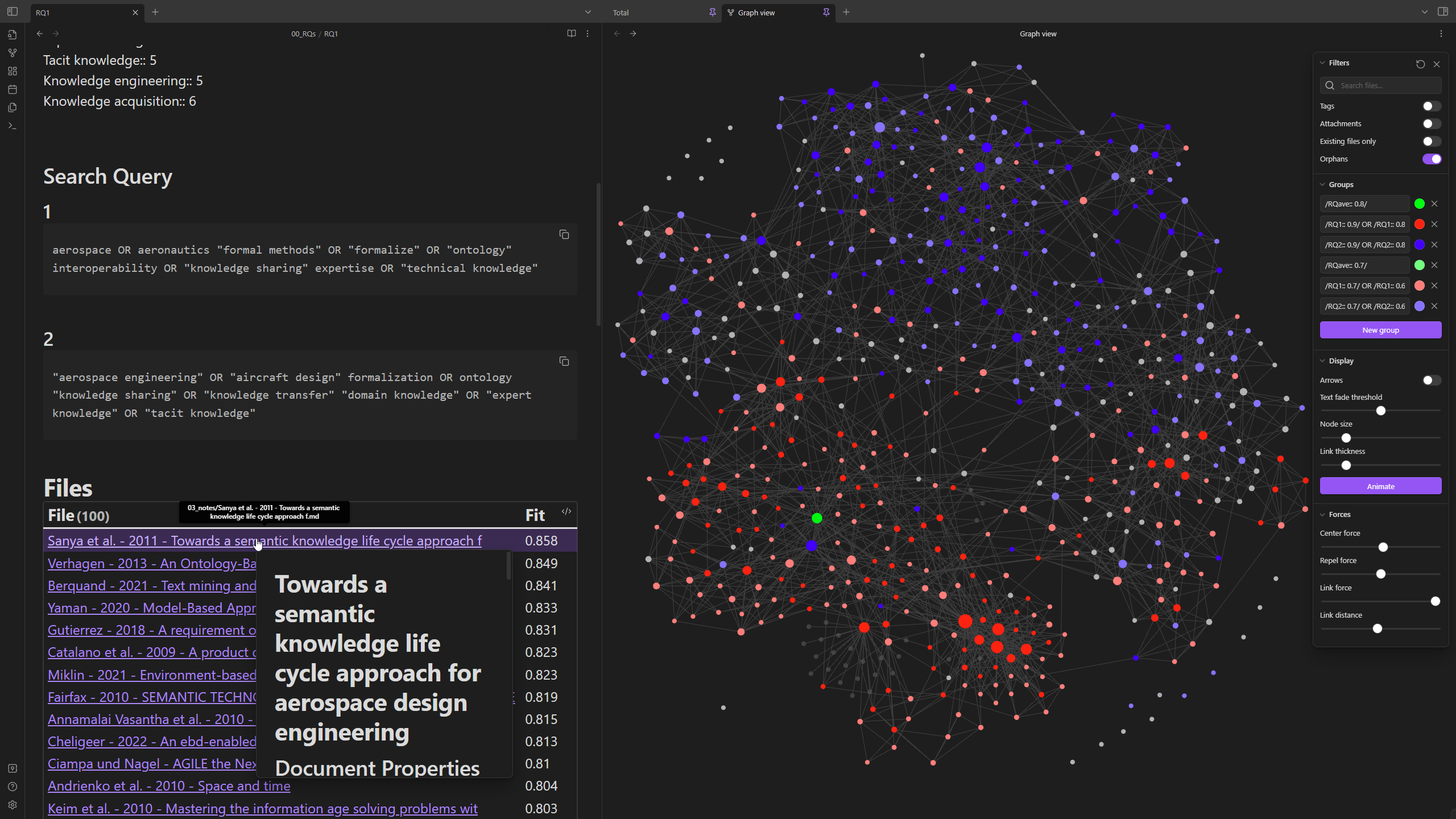}
    \caption{Obsidian overview of the analyzed library. The graph represents documents as nodes connected to similar documents and colored based on their relevance to a research question (RQ) (here: RQ1 red, RQ2 blue, average green). The RQ1 file on the left lists its dynamically ranked most relevant documents.}
    \label{fig:task6a}
    \vspace{-0.45cm}
\end{figure}

\paragraph{Notebooks.\label{par:script}}
Four Jupyter Notebooks guide through the workflow as depicted in \secref{fig:workflow}, each containing the markdown guideline, tool support, and code for the respective stage in the SWARM-SLR workflow.\footref{note:github-SWARM-SLR}
They are optimized for GitHub and can be used as an online reference with no prior technical expertise required.
Using checklists, visualizations, and step-by-step instructions, they assist the user in navigating through each requirement and completing each individual task of an SLR.
\secref{fig:task6a} illustrates the visualization during data discovery while showing the machine-readable representation of a research question with weighed keywords, queries, and dynamically generated lists of relevant documents.

\begin{figure}[tb]
    \centering
    \includegraphics[width=\textwidth]{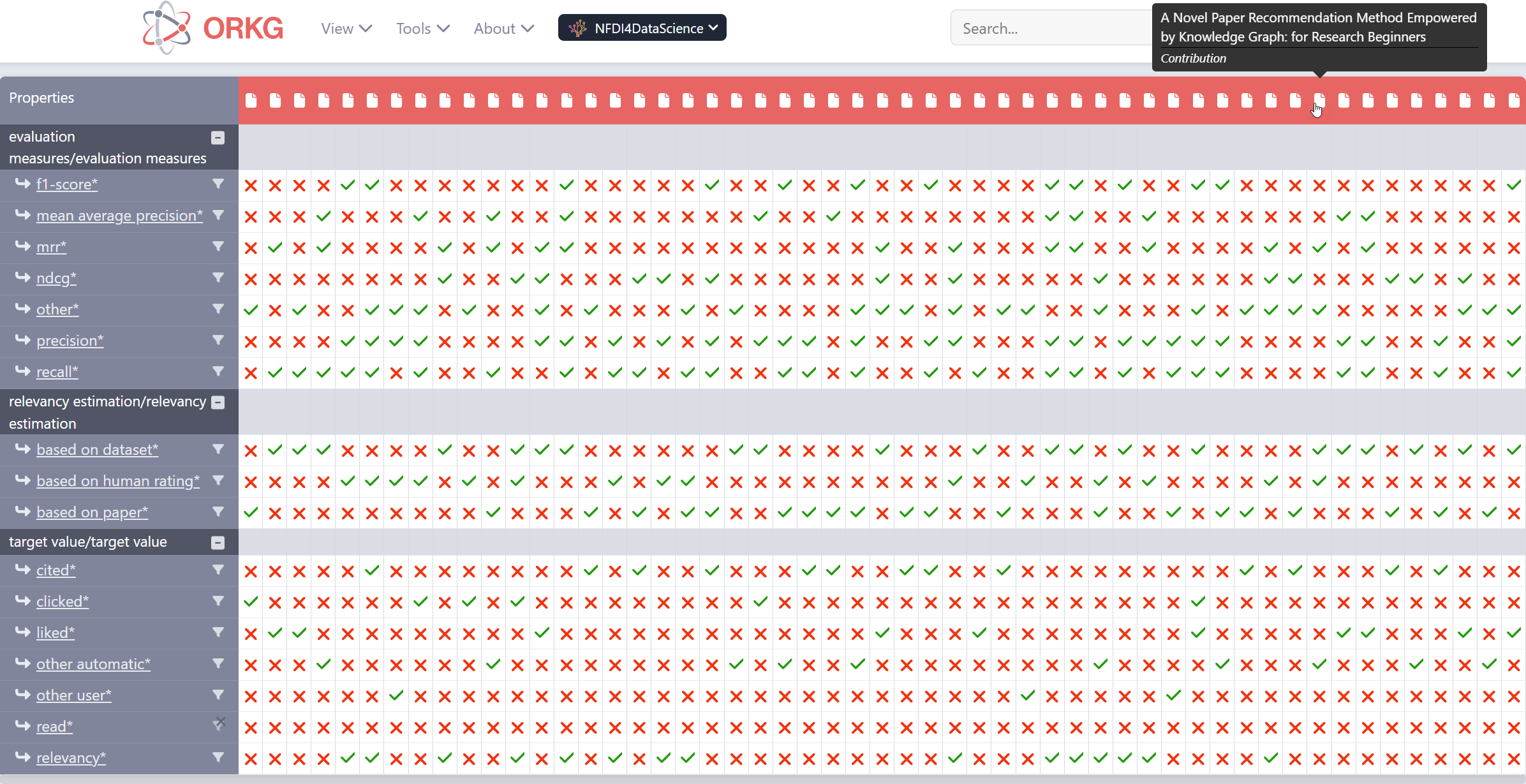}
    \caption{Scientific paper recommendation systems literature review by Kreutz and Schenkel~\cite{kreutz_scientific_2022}, represented in the ORKG, visualized as an ORKG comparison\footref{note:comparison}.}
    \label{fig:task71}
    \vspace{-0.45cm}
\end{figure}

Similarly, the literature further processed by Task 6 and 7.1 eventually create ORKG representations, which both directly contribute to the current SWARM-SLR in the form of comparisons as shown in \secref{fig:task71}, as well as to future SLR due to the persistent FAIR representation.
Taking the recommendation systems SLR by Kreutz and Schenkel~\cite{kreutz_scientific_2022} as an example, a template\footnote{Recommendation system template: \url{https://orkg.org/template/R673347/}} allows the transition from a table bound to two pages in a PDF publication to an interoperable knowledge graph. 
By implementing a subset of the Shapes Constraint Language (SHACL)~\cite{hussein_increasing_2023}, representations like these enable interoperable reuse, comparison\footnote{\label{note:comparison}Recommendation systems comparison: \url{https://orkg.org/comparison/R674188/}}, and future expansion. 
This results in dynamic reviews founded on connected knowledge and literature management, supporting asynchronous collaboration beyond the scope of single, disjoint articles~\cite{karras2023divide}.

\vspace{-0.25cm}
\section{Evaluation\label{sec:evaluation}}
This section presents the survey \nameref{sec:design} and \nameref{sec:results} to evaluate SWARM-SLR.

\subsection{Design\label{sec:design}}
Based on the validation guidance by Robson and McCartan \cite{robson_real_2016}, we designed a tandem-survey to validate SWARM-SLR:
One survey\footnote{\label{note:req}Requirement review survey: \url{https://survey.uni-hannover.de/index.php/555283}} validates which requirements correctly represent SLR tool requirements, the other\footnote{\label{note:tool}Tool assessment survey: \url{https://survey.uni-hannover.de/index.php/628237}} used these requirements to gauge the fitness of a given tool to a given task.

The modular structure provided through Rupp~\cite{rupp_requirements-engineering_2014} made the formulation of survey questions simple:
from ``\textit{Requirement: The tool should be able to help the user to efficiently <requirement> for the <result>.}'', we derived the statement template ``\textit{To complete the step `<step>', a tool should be able to help the user to efficiently <requirement> for the <result>.}''.
For the tool survey, ``\textit{a tool}'' was replaced with ``\textit{the tool}'', with a tool selection at the very beginning of the survey.
The respondents indicated their agreement with each statement on a 9-point Likert scale from "strongly agree" (A1) to "strongly disagree" (A9), with the option to not answer (A10). 
All steps of the same task were listed on the same page, headlined by the task.
These eight pages were framed by an opening page providing context, a page for demographic and literature survey expertise data, a freeform feedback question, and a closing page pointing towards the respective other survey.

The links to the surveys were distributed in mid-2023 and early 2024, iterating on the design of the workflow in between. 
In 2023, the surveys were shared directly with over 200 fellow researchers, primarily from the computer science and engineering domains.
Additionally, the surveys were distributed over outreach channels such as LinkedIn and SurveyCircle.
In 2024, the link to the tool survey was shared again with the 200 researchers, now with the added information where the SWARM-SLR workflow can be found and how it can be evaluated by selecting "other", then "SWARM-SLR" on the tool selection page.
For a long-term sustainable evaluation of these requirements and tools like Bosman and Kramer displayed, we have built an evaluation workflow around it, allowing for the continuous instream of differentiated evaluations, including new tool suggestions, added requirements and comments.\footnote{\label{foot:evaluation}\url{https://github.com/borgnetzwerk/tools/blob/main/scripts/SWARM-SLR/data/evaluation.ipynb}, \href{https://github.com/borgnetzwerk/tools/blob/main/scripts/SWARM-SLR/data//analysis555283.md}{/analysis555283.md} and \href{https://github.com/borgnetzwerk/tools/blob/main/scripts/SWARM-SLR/data/analysis628237.md}{/analysis628237.md}} All data processing, details and results are accessible there.

\vspace{-0.25cm}
\subsection{Results\label{sec:results}}
The required expertise to provide a valuable contribution to the topic in question proved to be a significant obstacle.
When faced with the depth of requirements, many respondents highlighted that the different tasks require very different domain knowledge.
Stage I appears to lean towards science management, stage II and III towards information retrieval, and stage IV towards scholarly communication.
In total, the requirement assessment was 
completed by 11 respondents, all of which were valid responses.
The tool assessment was 
completed by 37 respondents, of which 36 were valid.
Overall 12 tools have been evaluated, 6 of them by multiple respondents.
The SWARM-SLR Workflow has been evaluated by 11 respondents, for 40 minutes on average ($max = 60, min = 13, \sigma = 17.63$).
Two of these were from the engineering domain and nine were from computer science.

\vspace{-0.25cm}
\subsubsection{Requirements.\label{sec:req_eval}}
\begin{figure}[tb]
    \vspace{-0.45cm}
    \centering
    \includegraphics[width=\textwidth]{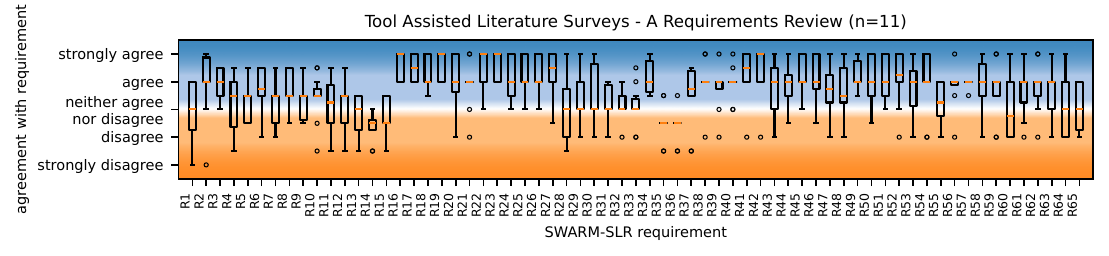}
    \caption{Boxplot of the agreement with the 65 requirements, as reported in the Requirements Review survey.}
    \label{fig:eval_req}
    \vspace{-0.45cm}
\end{figure}

\secref{fig:eval_req} displays the plotted agreement over all 65 requirements. The general agreement is rather positive, with selected exceptions like the re-evaluation with domain-experts (R12-R15, \secref{table:req1}) or the tf-idf and word-/document-embeddings (R35-R36, \secref{table:req2}).

When looking at expertise reported in the demographics\footref{foot:evaluation}, non-experts were extremely skeptical of tools supporting a re-evaluation with domain experts (R12-R15), while reported experts judged neutrally reserved and with balanced agreement.
A particularly interesting finding is that calculating tf-idf and word-/document-embeddings (R35-R36) was deemed overall not required.
Being probably the most technical requirements in the entire set, it is possible that these requirements are too abstract to be interpretable by general survey participants.
This assumption is particularly reinforced when observing that requirement R40 ``calculate document similarity'' is agreed on, while both \textit{tf-idf} and word embeddings are components of state-of-the-art document similarity analysis.

Overall, this evaluation identifies reserved skepticism towards a tool supporting \textit{Step 2.5 Reevaluate with domain experts} (R12-R15).
The vast majority of requirements, particularly in stages II through IV (R16-R65), are sustained and appear to properly reflect what tool support is required for SWARM-SLR.

\vspace{-0.25cm}
\subsubsection{Tools.\label{sec:tool_eval}}
\begin{figure}[tb]
    \vspace{-0.55cm}
    \centering
    \includegraphics[width=\textwidth]{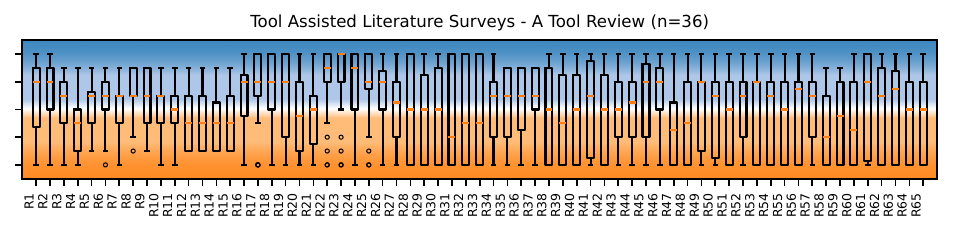}
    
    \vspace{-0.2cm}
    
    \includegraphics[width=\textwidth]{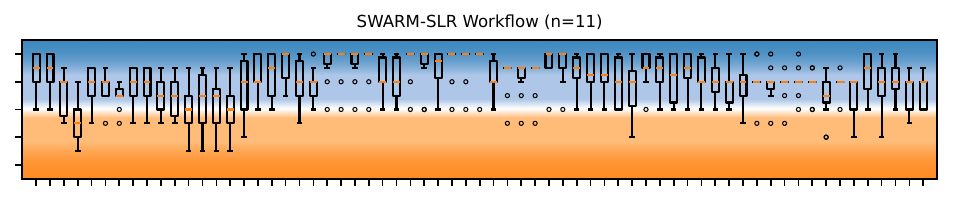}
    
    \vspace{-0.2cm}
    
    \includegraphics[width=\textwidth]{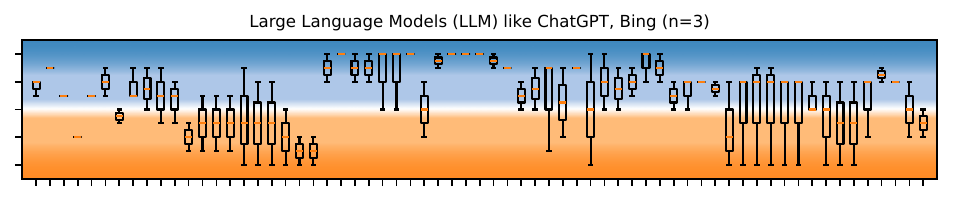}
    
    \vspace{-0.2cm}
    
    \includegraphics[width=\textwidth]{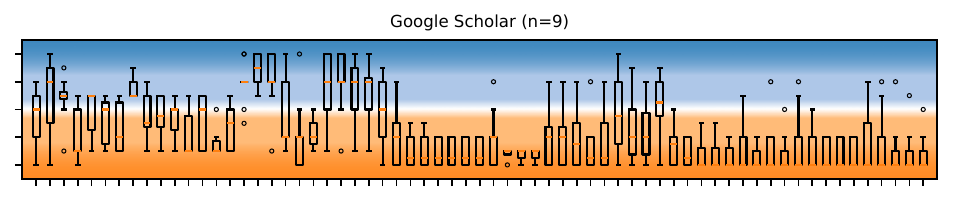}
    
    \caption{Boxplots of the agreement with the tools (All, SWARM-SLR Workflow, LLM, Google Scholar) supporting the 65 SWARM-SLR requirements. Identical axis labels to \secref{fig:eval_req}, reduced for better viewability.}
    \label{fig:tools_and_workflow}
    \vspace{-0.45cm}
\end{figure}

To interpret \secref{fig:tools_and_workflow}, we can set the highest values reported in the tool review as an upper boundary currently perceived possible.
This value may be biased due to subjective interpretation of the requirements, the tools' capability, or similar inflation.
The more respondents report for a tool, the clearer the picture, and the easier it is to identify outliers.
Google Scholar, for example, has been evaluated 10 times and clearly reflects its strength in search-related tasks.
LLMs similarly have been assessed 3 times, showing clear strengths in a wide range of tasks, while simultaneously showing strong limitations and skepticism in many of them.
Finally, the SWARM-SLR workflow has been more thoroughly evaluated than all other tools combined (total: 442 min, median: 39 min vs. total: 386 min, median: 10 min).
It was able to streamline the workflow while apparently addressing most of the requirements, with some reservations at \textit{Step 2.5 Reevaluate with
domain experts} (R12-R15) as well as in \textit{Step 7.3 Critique the literature} (R58).

Outside of the requirements survey, the SWARM-SLR workflow tool evaluation was the only one that included comments.
Noteworthy is a positive sentiment (``\textit{Can't wait to test SWARM on my next literature review. Seems like a thoroughly thought-through methodology. Cheers to the development!}'', ``\textit{The approach is structured in a nice way and has a comprehensive list of tasks. Which is good to have a detailed step-by-step guide to do such a complex task as systematic reviews, especially for scientists, [...]}''), while some still highlight the road ahead (``\textit{[...], but it is still daunting. For instance, the "Extract structured data from documents" in "Select" is great in encompassing all the small tasks in code which make it easy to the user/researcher. While "Analysis" and "Synthesis" tasks are not as supported}'').

\section{Discussion and Future Work\label{sec:discussion}}
Much like Bosman and Kramer's work~\cite{noauthor_400_tools_and_innovations_in_scholarly_communication_nodate,bosman_innovations_2016}, the efforts of this contribution do not end with the publication.
The tandem-survey\footref{note:req}$^{,}$\footref{note:tool} and evaluation remains open for the foreseeable future, encouraging SRA experts from all stages to contribute their expertise for future data collection and integration into the SWARM-SLR framework.
Detailing and aligning the different abstraction levels of tasks, steps, results and requirements along the SWARM-SLR workflow requires further interdisciplinary collaboration and evaluation, particularly \textit{Task I: Planning
a review}.
With additional experts evaluating requirements, providing tool assessments and feedback on the workflow, we expect the mapping of accessible and interoperable tools to be achievable and beneficial to scholarly communication and knowledge distribution.

The current state provides a ready-to-use system for all three major contributions:
The 65 \textit{SWARM-SLR Requirements} are a starting point to which none of our survey participants had any further requirements to add, but foreseeable many SLR experts could further strengthen this set.
Overall 12 \textit{SWARM-SLR Tools} have been evaluated, only half of them by multiple respondents, and evidently, hundreds more are available for systematic evaluation, a process that could also be further improved by implementing measurable and quantifiable criteria beyond Likert-scale agreement.
It has been shown that the \textit{SWARM-SLR Workflow} already fulfills most of the requirements by connecting the tools to guide and support the users.
Once the tool catalog and requirement set grow further, this workflow similarly needs to be further adapted.
The current workflow system, while capable, already has some room for improvement, as evidenced by the continuous improvement during the ongoing Aerospace Knowledge Engineering SWARM-SLR use-case.

If this endeavor is continued, we see the potential for eventually\break{}(semi-)automating the SLR workflow. 
Given a common requirement set like the one presented in this work is established, against which SRA tools are systematically evaluated, the SWARM-SLR tool catalog should grow to provide solutions to most tasks, as already evident.
If these tools are made interoperable in a SWARM-SLR workflow as demonstrated, while further tool development is aimed at evidently less supported tasks, this method should produce a collaboratively developed and improved workflow.
This common workflow, potentially specialized to fit certain domains, but unified by the utilization and production of FAIR data artifacts, should lead to domain-independently raising researchers' capabilities to review and manage the dramatically growing scholarly knowledge corpus.

\section{Conclusion\label{sec:conclusion}}
In this article, we proposed the Streamlined Workflow Automation for Machine-actionable Systematic Literature Reviews (SWARM-SLR).
This includes the requirements, tools measured against these requirements, and a workflow synthesizing these.
The requirements were elicited from different domains, streamlining their perspectives and guidelines into a set of 65 requirements.
These requirements were evaluated through a tandem-survey alongside potential tools and eventually the implementation of the SWARM-SLR workflow, utilizing a selection of these tools along the research life-cycle from ``Formulate research interest'' to ``Write the literature review''.
Evidently, these requirements already provide a comprehensive framework to situate a tool's strengths and weaknesses over the SWARM-SLR workflow.
We advise to start assessing tools in their usage and development with the SWARM-SLR framework, as well as utilizing these tools and workflow for an efficient AI-supported literature review process.

This initial version of the SWARM-SLR workflow already displays the potential to improve literature-bound knowledge synthesis.
The tools evaluated also display specific gaps that can be filled by crowdsourced tool mapping and development.
Evidently, all phases, particularly search and retrieval, information extraction, knowledge synthesis, and distribution, can already greatly profit from these advancements.
We advise furthering these efforts to develop interfaces between these tools and eventually shape an accessible workflow completely covering all stages.
As a result, we envision SWARM-SLR to significantly contribute to digitizing scholarly communication, empower scientists to master the publication flood, to significantly increase AI assistance, and to overall make surveying research more FAIR, collaborative, and effective.

\section*{Acknowledgements}
This work was co-funded by the DFG SE2A Excellence Cluster, as well as the NFDI4Ing project funded by the German Research Foundation (project number 442146713) and NFDI4DataScience (project number 460234259).

\bibliographystyle{splncs04}
\bibliography{main}

\end{document}